\documentclass[conference]{IEEEtran}
\IEEEoverridecommandlockouts
\usepackage{braket}
\usepackage[dvipsnames]{xcolor}
\usepackage{url}
\usepackage{musicography}
\usepackage[inline, draft, nomargin]{fixme}
\fxsetup{theme=color}
\FXRegisterAuthor{xl}{atenv}{XL}
\FXRegisterAuthor{ab}{abenv}{\color{Rhodamine}AB}
\FXRegisterAuthor{rt}{rtenv}{\color{blue}RT}
\usepackage{graphicx, amsmath, amsthm, amssymb}
\usepackage{fullpage}
\usepackage{caption}
\usepackage[english]{babel}
\usepackage{amssymb,hyperref}
\usepackage{amsmath}
\usepackage{array, fancyhdr, bm, listings}
\usepackage{algorithmic,algorithm}
\usepackage{subcaption}
\usepackage{thmtools, thm-restate}
\usepackage{fancyhdr,mathtools}
\renewcommand{\baselinestretch}{1.10}

\newfont{\bssten}{cmssbx10}
\newfont{\bssnine}{cmssbx10 scaled 900}
\newfont{\bssdoz}{cmssbx10 scaled 1200}

\newcounter{topic} 

\definecolor{gray}{rgb}{0.5,0.5,0.5}

\makeatletter
\newenvironment{breakablealgorithm}
  {
   \begin{center}
     \refstepcounter{algorithm}
     \hrule height.8pt depth0pt \kern2pt
     \renewcommand{\caption}[2][\relax]{
       {\raggedright\textbf{\ALG@name~\thealgorithm} ##2\par}%
       \ifx\relax##1\relax 
         \addcontentsline{loa}{algorithm}{\protect\numberline{\thealgorithm}##2}%
       \else 
         \addcontentsline{loa}{algorithm}{\protect\numberline{\thealgorithm}##1}%
       \fi
       \kern2pt\hrule\kern2pt
     }
  }{
     \kern2pt\hrule\relax
   \end{center}
  }
\makeatother
\def\BibTeX{{\rm B\kern-.05em{\sc i\kern-.025em b}\kern-.08em
    T\kern-.1667em\lower.7ex\hbox{E}\kern-.125emX}}
\begin{document}

\newcommand\xin[1]{{\color{cyan}{Xin: #1}}}
\title{Locality-aware Qubit Routing for the Grid Architecture \\
}

\author{\IEEEauthorblockN{Avah Banerjee}
\IEEEauthorblockA{
 \textit{Dept. of Computer Science} \\
\textit{Missouri S\&T}\\
}
\and
\IEEEauthorblockN{Xin Liang}
\IEEEauthorblockA{
 \textit{Dept. of Computer Science} \\
\textit{Missouri S\&T}\\
}
\and
\IEEEauthorblockN{R. Tohid}
\IEEEauthorblockA{
 \textit{Center for Computation and Technology} \\
\textit{Louisiana State University}\\
}

}

\maketitle

\begin{abstract}
Due to the short decohorence time of qubits available in the NISQ-era, it is essential to pack (minimize the size and or the depth of) a logical quantum circuit as efficiently as possible given a sparsely coupled physical architecture.
In this work we introduce a locality-aware qubit routing algorithm based on a graph theoretic framework. 
Our algorithm is designed for the grid and certain ``grid-like" architectures. We experimentally show the competitiveness of algorithm by comparing it against the approximate token swapping algorithm, which is used as a primitive in many state-of-the-art quantum transpilers. Our algorithm produces circuits of comparable depth (better on random permutations) while being an order of magnitude faster than a typical implementation of the approximate token swapping algorithm.
\end{abstract}


\vspace{-1em}
\section{Introduction}
\underline{N}oisy \underline{I}ntermediate \underline{S}cale \underline{Q}uantum (NISQ) - era quantum computers are constrained by various hardware limitations.
The underlying technology (for example, superconducting qubits, trapped ion etc.) determines error rates and realizability of different single and two qubit gate operations. The small number of physical qubits available to NISQ processors \footnote{as of writing this paper the number of qubits on available systems range from $5$ to about $200$} limits the use of quantum error correcting codes; a feature to be expected for fault tolerant quantum computers.

In the meantime various engineering as well as algorithmic solutions has been proposed to reduce the overall circuit error by carefully navigating the constraints imposed by the hardware.  
One such constraint, which particularly manifests in devices based on the superconducting qubit architecture, limits the set of pairs of physical qubits that can take part in a two qubit gate operation.
The pairs of physical qubits which can take part in a two qubit gate operation are said to be \emph{coupled}. 
Suppose $Q_L$ is a logical quantum circuit that we wish to execute on a given hardware.
We assume that not all pairs of physical qubits are coupled.
In this case we need to map the logical qubits to physical qubits \footnote{Note that due to the absence of any usable error correcting codes in the NISQ era, these mappings are one to one.}. 
This mapping must ensure that every pair of logical qubits that take part in a two qubit gate is mapped to a pair of physical qubits that are coupled. 
However, in most cases, there is no single mapping that can simultaneously satisfy all of the coupling requirements imposed by $Q_L$. 
In such a situation, logical qubits are remapped, possibly multiple times, to different physical locations (physical qubits) so that all the two qubit gates in $Q_L$ are executed on a schedule satisfying the dependencies in $Q_L$. A single qubit gate can be executed in-place, without moving the qubits. Hence, for clarity of exposition we can ignore the presence of single qubit gates in $Q_L$ when discussing qubit routing. However, in practice the scheduling of two qubit gates does depend on single qubit gates and hence plays a role in determining the  depth of the physical circuit ($Q_P$).

If a qubit is remapped, it has to be physically moved to its new location. This step is called routing and is usually achieved by adding appropriate swap gates  to the logical circuit. A swap gate exchanges the state of its  two input qubits. In some hardware, a swap gate is constructed using a sequence of three controlled-not gate.
However these extra swap operations increase the size (the number of gates) and the depth of the circuit (the length of the critical path in the circuit).
Because the transformed circuit may then be too big to be reliably implemented on the given hardware, the output state of $Q_P$ may significantly deviate from its expected state (the output state of $Q_L$).
If the output state is classical (result of some measurements), we may be able to mitigate the problem by executing $Q_P$ multiple times. 
However, such a strategy invariably leads to more resource utilization.

As such, it is important to ``pack" the logical circuit within a physical circuit of small depth by optimizing the mapping and the routing steps. 
In this paper, we focus on optimizing routing of qubits for the grid and ``grid-like" architectures. 
Almost all superconducting qubit based architectures are planar. That is, the coupling of the qubit pairs can be represented by some planar graph. Majority of these planar architectures are ``close to" some grid graph.
This was our main motivation for studying routing on this type of architectures. 

Specifically, we design a routing algorithm for the grid by exploiting the locality in the underlying permutation. 
Our algorithm leads to a significantly better performance than and produces routing of  depth comparable to the state of the art. Our algorithm can be extended to graphs which are Cartesian product of two graphs.
Our algorithm builds upon the routing via matching framework introduced by Alon et. al. \cite{alon1994routing}.
As such, it is a parallel routing scheme as opposed to the token swapping framework commonly used. 
It is expected to benefit a wide range of quantum programs including simulation of spatially local Hamiltonians.

\begin{figure*}[htbp]
  \centering
  \includegraphics[width=16cm]{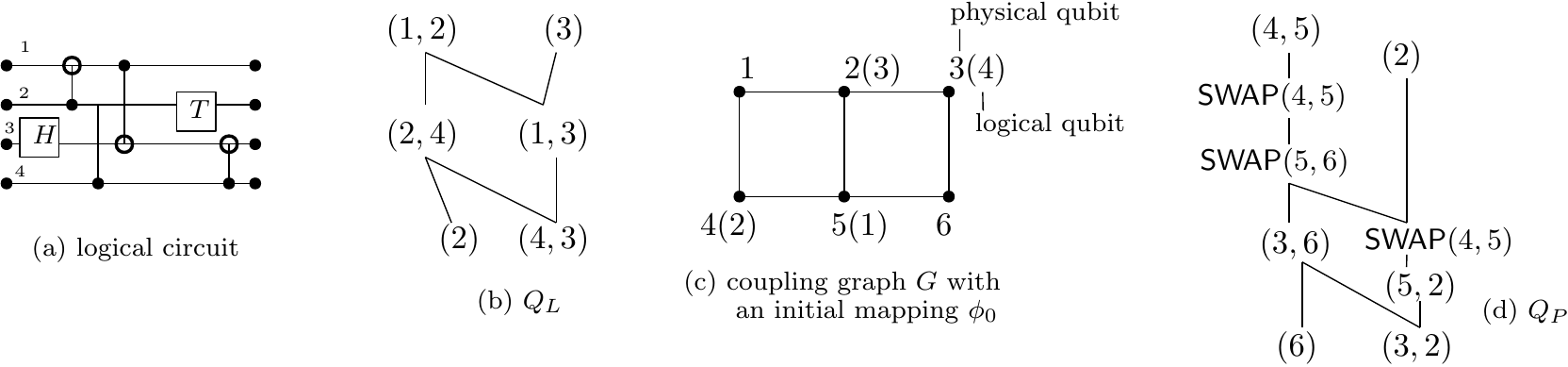}
  \caption{An example of routing to make the logical circuit in (a)  conform to the physical couplings according to (c).}
  \label{fig: routing example}
\end{figure*}
\section{Problem Formulation}
In this section, we formally introduce the qubit routing problem and the routing via matching framework.
An example is given in Figure \ref{fig: routing example}.
Physical couplings between the qubits can be represented by an undirected simple graph, usually referred to as the \emph{coupling graph}.
We will use $G = (V,E)$ to denote this graph (see Figure \ref{fig: routing example}-(c)).
In this paper we assume $G$ to be the $m \times n$ grid graph.
A vertex in $V$ is identified with a pair of indices $(i,j)$ on the grid ($i \in [m]$ and $j \in [n]$\footnote{$[n] = \{1,\ldots,n\}$}). 
Figure \ref{fig: routing example}-(a) gives an example of a logical circuit with four qubits and five gates. 
In Figure \ref{fig: routing example}-(b) this circuit is represented as a directed acyclic graph ($Q_L$).
The vertices of $Q_L$ correspond to the gates of the circuit and the edges represent the dependencies among them.
The label(s) on the vertices correspond to the qubit(s) involved in the gate.
Figure \ref{fig: routing example}-(d) gives a possible physical realization $Q_P$ of $Q_L$ on the coupling graph $G$. 
The circuit $Q_P$ is \emph{feasible} for $G$ as all its gates use qubits that are adjacent in $G$.
We see that both the size ($5 \to 9$) and the depth ($3 \to 6$) of $Q_P$ is greater than that of $Q_L$. 
These increases in size and depth invariably make it more likely that the output of $Q_P$ will deviate significantly from that of $Q_L$, which is particularly true for NISQ devices without error correction.
The goal of the transformation algorithm, the \emph{transpiler}, is to produce a feasible circuit for a given coupling graph, which is pareto-optimal with respect to the objectives of minimizing the physical circuit size and depth.
Note that a unique solution that minimizes both the size and depth of $Q_P$ may not exist.
Unfortunately, this problem is $\mathsf{NP}$-hard, even if we want to optimize one of the objectives.
Further, seeking optimally may not even be of much use if the optimal circuit is not that far from (in terms of size and/ or depth) from some arbitrary feasible circuit.
This is particularly the case when $G$ is quite sparse and $Q_L$ has many infeasible gates. As an extreme example, suppose $Q_L$ be the $\mathsf{QFT}$ circuit on $n$-qubits and $G=P_n$ is the path with $n$ vertices. It is an easy exercise to see that per layer of the logical $\mathsf{QFT}$ circuit we need $\Omega(n)$ $\mathsf{SWAP}$ gates.

To make the above optimization problem feasible, it is often decomposed into an alternating sequence of \emph{mapping} and \emph{routing} problems. 
In the mapping phase, we try to pick a mapping of the logical qubit to the physical qubit. 
For example,  Figure \ref{fig: routing example}-(c) shows an initial mapping of the logical qubits to the vertices of $G$.
In the routing phase we move the logical qubits to their new locations determined by the mapping. 
In this paper we focus on the latter.
To this end, our routing algorithm can be used in any transpiler that uses the above framework as an alternative to the routing algorithm used there.

The destinations of the logical qubits in the routing phase is given by a permutation on $V$.
Oftentimes, we do not care about the location of some qubits.
In such a case, the destinations are given by a bijection $f: S \to R$, where $S, R \subset V$. We can extend $f$  to a permutation by selecting destinations for the don't-care qubits. Here we assume this extension has already been determined by the transpiler and we are given a permutation to route.
In the routing via matchings model, the routing schedule is determined by a sequence of matchings in $G$.
We move the logical qubits along the edges in these matchings.
More specifically, for each edge $(i,j)$ in a matching we  add a $\mathsf{SWAP}_{i,j}$ gate to the circuit with physical qubits $i$ and $j$ as inputs.
Hence a matching corresponds to a layer of a mutually disjoint set of $\mathsf{SWAP}$ gates which can be executed in parallel.
The depth of the circuit is increased by the number of matchings in the routing schedule.
Therefore, our goal is to identify a sequence of matchings that minimizes the depth.
In addition, the computation should be efficient and scalable for the scheme to work in practice.
Unfortunately, computing an optimal matching sequence is $\mathsf{NP}$-hard\cite{banerjee2017new}.
As of yet there is no approximation guarantee for this problem, except for the case when $G$ is the path graph.
In contrast, for the serial variant of the problem, where we only care about minimizing the number of swaps, the approximate token swapping algorithm by Miltzow et. al. \cite{miltzow2016approximation} has an approximation factor of 4.
Interestingly, the swaps discovered by the token swapping algorithm produces a routing schedule with depth comparable to our parallel routing algorithm.

\section{Related Work}
There have been a considerable number of recent studies on the qubit mapping problem (\cite{murali2020software,sivarajah2020t,li2019tackling,siraichi2018qubit}). Some of these methods combine mapping and routing to one combinatorial optimization problem (example \cite{murali2019noise}) or using routing time as a measure of efficacy of the mapping scheme (example \cite{childs2019circuit}). 
In contrast, only a handful of work is proposed to specifically deal with the qubit routing problem in isolation, when a mapping is already determined. In this section we briefly go over the literature on qubit routing.

Token swapping either in the serial or in the parallel setting (a.k.a routing via matchings) has been studied for close to three decades. Some relevant results can be found in (\cite{banerjee2017new,alon1994routing,miltzow2016approximation,yamanaka2015swapping}) and the references therein.
Here we briefly mention some work relevant to routing qubits that has been proposed in the last few years.
Childs et. al. \cite{childs2019circuit} initiated a systematic study of various routing (as well as qubit mapping) strategies for both general as well as special classes of coupling graphs. 
The (partial) routing algorithms proposed there mostly used standard methods from earlier works by Alon, Miltzow and others \cite{alon1994routing,miltzow2016approximation}. Routing via reversals has also been applied in the qubit routing setting. 
This is a particularly promising approach as the reversal of $n$ qubits along a line can be carried out faster using certain topological transformations of spin chains \cite{bapat2021quantum} in the Majorana picture. Such schemes have been well studied for linear networks (as reversal of spin chains in condensed matter physics - for example in \cite{albanese2004mirror,karbach2005spin} etc. and more recently in \cite{bapat2022nearly}).
Bopat et. al. \cite{bapat2021quantum} proposed a qubit routing scheme for general graphs by reducing the problem to that of routing on a tree.                                                 

\begin{figure*}[htbp]
  \centering
  \includegraphics[width=16cm]{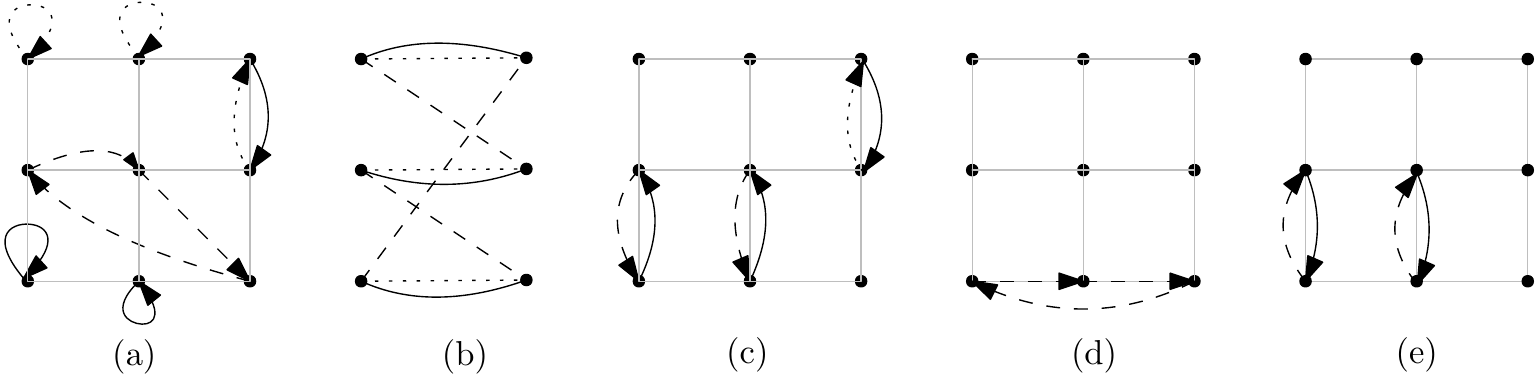}
  \caption{An example of routing on a $3 \times 3$ grid. (a) Arrows indicate the destination of the qubits. (b) Shows the bipartite multi-graph $G^[1,3]$ indicating qubit movements between columns. Edges that are part of different perfect matchings are distinguished using different styles (solid, dashed and dotted). (c)-(e) are the three rounds of the routing. For example the qubit initially at $(2,2)$ moves to $(3,2)$ and then to $(3,3)$ after the end of the second round. Note that each round may involve multiple steps, where each step is a set of concurrent swap operations.}
  \label{fig: grid route}
\end{figure*}                                              

\section{The Proposed Algorithm for Grid}
In this section, we present our qubit routing algorithm for the grid graph.
The algorithm builds on the 3-step grid routing algorithm in~\cite{alon1994routing}.
Just like the algorithm in \cite{alon1994routing} ours will also work on any graph $G$ which can be expressed as a Cartesian product $G_1 \square G_2$ of two graphs $G_1, G_2$. Vertices of $G$ are ordered pairs $(u,v)$ where $u \in G_1$ and $v \in G_2$. There is an edge between two vertices $(u,v)$ and $(u',v')$ if and only if either $(u,u')$ is an edge of $G_1$ or $(v,v')$ is an edge of $G_2$. The $m \times n$ grid graph is the Cartesian product of $P_m \square P_n$, where $P_n$ is the path with $n$ vertices. 
In what follows we present our algorithm on the grid graph.
After that, we will briefly discuss the modifications needed to extend it to Cartesian product graphs at the end of this section.

We begin by briefly discussing the original grid routing algorithm of~\cite{alon1994routing}. An example is shown in Figure \ref{fig: grid route}. Let $G$ be an $m \times n$ grid graph.
Suppose the permutation $\pi$ on $G$ sends some qubit at location $(i,j)$ to $(i',j')$.
For a fixed $j'$ there are exactly $n$ qubits that will be sent to the column labeled $j'$.
By successive applications of Hall's marriage theorem, we can identify a set of $n$ permutations ($\sigma_1,\ldots,\sigma_n$) on the columns with the following property.
 After routing the qubits in column $i$ using $\sigma_i$, the destination columns  of every qubit will be unique in each row. That is, we can route the qubits along the rows in parallel so that after we are done with this round, every qubit is in its correct destination column.
Then in the next round, we route the qubits in each column in parallel.
As such, this algorithm involves three rounds of routing in a column-row-column order. We will denote this routing scheme as $\mathsf{GridRoute}(G,\pi ;\sigma_1,\ldots,\sigma_n)$, which returns a sequence of matchings $(M_1,\ldots,M_t)$ of $G$.
However, we can also perform the routing in the row-column-row order ($\mathsf{GridRoute}(G^T,\pi^T;\sigma_1,\ldots,\sigma_m)$\footnote{Here, $G^T$ is the transpose of the grid $G$ (determined by the automorphism which sends $(i,j) \to (j,i)$) and $\pi(i,j) = (i',j')$ iff $\pi^T(j,i) = (j',i')$ }) and finally choose the strategy that leads to the smallest depth.
In each round the parallel routings along the rows or the columns is done using the odd-even transposition algorithm for routing on a path.
The above three-round strategy can be extended to the case when $G = G_1\square G_2$ as follows.
$G$ can be thought of as a ``grid-like" graph where each row (resp. column) is replaced by copy of $G_1$ (resp. $G_2$).
In each round we route the qubits in parallel on the respective copies of $G_1$ (resp. $G_2$)  using some appropriate routing algorithms for $G_1$ (resp. $G_2$). 
In a similar manner, we can extend our locality aware routing algorithm for grids to this more general case.

\begin{figure}[htbp]
  \centering
  \includegraphics[width=0.3\columnwidth]{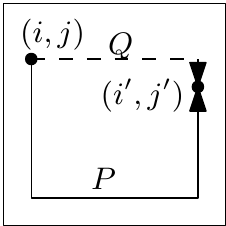}
  \caption{Suppose $\pi(i,j) = (i',j')$. Depending on the permutation chosen in the first round the qubit at $(i,j)$ may end up getting routed via the path $P$ instead of a shorter path $Q$.}
  \label{fig: bad path}
\end{figure}

The grid routing algorithm described above overlooks the possible locality in the underlying permutation, which exists in a wide range of quantum applications. 
More specifically, there are cycles of the permutation $\pi$ that are contained within small regions of the grid in many of these applications.
The permutations $(\sigma_1,\ldots, \sigma_m)$ are chosen by finding a set of $m$ perfect matchings on a bipartite multi-graph, which, unfortunately, are done in an arbitrary manner and may end up creating a schedule with unnecessary overhead (see for example Figure \ref{fig: bad path}). 
By considering the locality of qubit movement, our algorithm ensures that the permutations selected in the first stage does not make any qubit take a path to reach their destination that is too long relative to a path used in an optimal routing scheme. This will promise smaller depth in the transpiled circuit.

\subsection{Preliminaries}
Before proceeding to describe our algorithm, we introduce some additional notations and definitions.
We define a bipartite multi-graph $G^{[a,b]}([n],[n])$, where using $[n]$ we identify the set of $n$ columns  of $G$. For notational simplicity, we use $G^{[a,b]}$ to refer to this graph.
For each pair $((i,j),(i',j'))$ of vertices in $G$, where $i \in \{a,\ldots,b\}$, there is an edge labeled $(i,i')$ between the vertex labeled $j$ and $j'$ in $G^{[a,b]}$ iff $(i',j') = \pi(i,j)$. 
Figure \ref{fig: grid route}-(b) shows the graph $G^{[1,3]}$ corresponding to the permutation in (a). 
Let $M = \{(i_1,i_1'),\ldots, (i_n,i_n')\}$ be a perfect matching of $G^{[1,m]}$.
We define a metric $\Delta$ that we use to determine how far a matching is from some row in $G$.
\begin{align*}
    \Delta(M,r) = \sum_{j=1}^n|i_j-r|+ \sum_{j=1}^n|i'_j-r|
\end{align*}
Let ${\cal P}$ be a set of all perfect matchings of $G^{[1,m]}$ (see \cite{alon1994routing} for a proof of their existence).
We define a complete bipartite graph $H({\cal P},[m])$ where the left vertices are the matching in ${\cal P}$ and the right vertices are the rows of $G$. 
Lastly, we introduce the \emph{maximum cardinality bottleneck bipartite matching} ($\mathsf{MCBBM}$) problem (\cite{gabow1988algorithms,punnen1994improved}).
Given an edge weighted bipartite graph, the task in $\mathsf{MCBBM}$ is to find a maximum matching which minimizes the maximum weight of any edge in the matching.

\subsection{The Locality-aware Routing Algorithm}
\begin{breakablealgorithm} 
\caption{Main Procedure}
\begin{algorithmic}[1]
\REQUIRE A $m \times n$  grid graph $G$, a permutation $\pi$ 
\ENSURE A sequence of matchings $\cal M$ of $G$ 
\STATE $(M_1,\ldots, M_{t}) \leftarrow \mathsf{LocalGridRoute}(G, \pi)$
\STATE $(M'_1,\ldots, M'_{t'}) \leftarrow \mathsf{LocalGridRoute}(G^T, \pi^T)$
\IF{$t \le t'$} 
\STATE \textbf{return} $(M_1,\ldots, M_{t})$
\ELSE 
\STATE \textbf{return} $(M'_1,\ldots, M'_{t'})$
\ENDIF
\end{algorithmic}
\label{alg: main}
\end{breakablealgorithm}

\begin{breakablealgorithm} 
\caption{$\mathsf{LocalGridRoute}(G, \pi)$}
\begin{algorithmic}[1]
\REQUIRE A $m \times n$  grid graph $G$, a permutation $\pi$ 
\ENSURE A sequence of matchings $\cal M$ of $G$ 
\STATE ${\cal M} \leftarrow \emptyset$
\STATE \textbf{construct} $G^{[1,m]}$ \\ //\texttt{first we find a set of $m$ perfect matchings in $G^{[1,m]}$ }\\ //\texttt{let $E^c$ be the set of edges in $G^{[1,m]}$}
\STATE $w \leftarrow 0$ \qquad//\texttt{search window size}
\STATE ${\cal P}\leftarrow \emptyset$ \\ //\texttt{apply a doubling search}
\WHILE{$|{\cal P}| < m$}
\STATE $r \leftarrow 1$ \qquad//\texttt{starting row}
\FOR{$0 \le i \le \left\lfloor \frac{m}{w+1} \right\rfloor$}
\STATE Find all perfect matchings (if any) in
 $G^{[r,\min(r+w,m)]}$ and add them to ${\cal P}$ \\//\texttt{remove the edges in $\cal P$ from $G^{[1,m]}$}
\STATE $E^c \leftarrow E^c \setminus \cup_{M \in {\cal P}}M$
\STATE $r \leftarrow r+w+1$
\STATE $i \leftarrow i+1$
\ENDFOR
\IF{ $w = 0$ }
\STATE $w \leftarrow 1$
\ELSE 
\STATE $w \leftarrow 2w$
\ENDIF
\ENDWHILE
\STATE \textbf{construct} $H$  from ${\cal P}$
\STATE $M^{\sh} \leftarrow \mathsf{MCBBM}(H)$\\//\texttt{Using $M^{\sh}$ we identify a row in $G$ for each perfect matching in ${\cal P}$}\\//\texttt{construct the permutations $\sigma_1, \ldots , \sigma_n$}
\FORALL{$(i,i') \in M \in {\cal P}$}
\STATE $\sigma_j(i) \leftarrow r$\ //\texttt{where $(M,r) \in M^{\sh} $ and $\pi(i,j)=(i',j')$}
\ENDFOR
\STATE \textbf{return} $\mathsf{GridRoute}(G,\pi;\sigma_1,\ldots,\sigma_n)$
\end{algorithmic}
\end{breakablealgorithm}

\subsection{Correctness, Runtime Analysis and Extension}
\emph{Correctness.} $\mathsf{LocalGridRoute}(G, \pi)$ will eventually discover a set of $m$ perfect matchings.
It follows then that for a fixed $r \in [m]$, the set $\{j' \mid\ \pi(\sigma_j^{-1}(r),j) = (i',j')\}$ has $n$  elements.
Hence the permutations $(\sigma_1,\ldots,\sigma_n)$ satisfy the necessary requirements of the  $\mathsf{GridRoute}$ algorithm.

\emph{Running Time.}  The main \texttt{while} loop at line--5 runs at most $\lceil \log  m \rceil$ times. We can find a perfect matching (or determine there is none) in
 $G^{[a,b]}$ in time $O(mn\sqrt{n})$ \cite{kao1999decomposition}, since $G^{[1,m]}$ has $mn$ edges.
 Hence the main \texttt{while} loop takes $O(m^2n\sqrt{n})$ time per iteration and  $\widetilde{O}(m^2n\sqrt{n})$ time in total. Here $\widetilde{O}$ hides a poly-logarithmic factor in $m,n$. Since $H$ is a complete bipartite graph with $m$ vertices and ${m \choose 2}$ edges, using the algorithm of Punnen and Nair \cite{punnen1994improved}we can solve $\mathsf{MCBBM}$ on $H$ in $\widetilde{O}(m^{2.5})$ time, which is dominated by the previous bound.  The rest of the algorithm involves computing the actual swap sequence which takes time linear in the size of $G$. This cost is dominated by the work done before line 24. Hence the total time taken by $\mathsf{LocalGridRoute}(G, \pi)$ is $\widetilde{O}(m^2n\sqrt{n})$ and the main procedure (Algorithm \ref{alg: main}) takes $\widetilde{O}(m^2n\sqrt{n}+mn^2\sqrt{m})$ time.
 
 \emph{Extension to Cartesian Products.} 
 We can extend our algorithm for Cartesian product graphs by extending the $\mathsf{GridRoute}$ subroutine appropriately. Specifically, replacing the odd-even transposition with routing algorithms for $G_1$ and $G_2$.
 However, depending on the structure of $G_1, G_2$,  optimizing for locality may not be that significant.
 If $G_1, G_2$ are somewhat path-like in a technical sense (for example their \emph{path-widths} are small), then we expect our locality aware algorithm to produce useful improvements over the naive algorithm.
 
\section {Experimental Results}
\begin{figure}[htbp]

  \centering
  \includegraphics[width=\columnwidth]{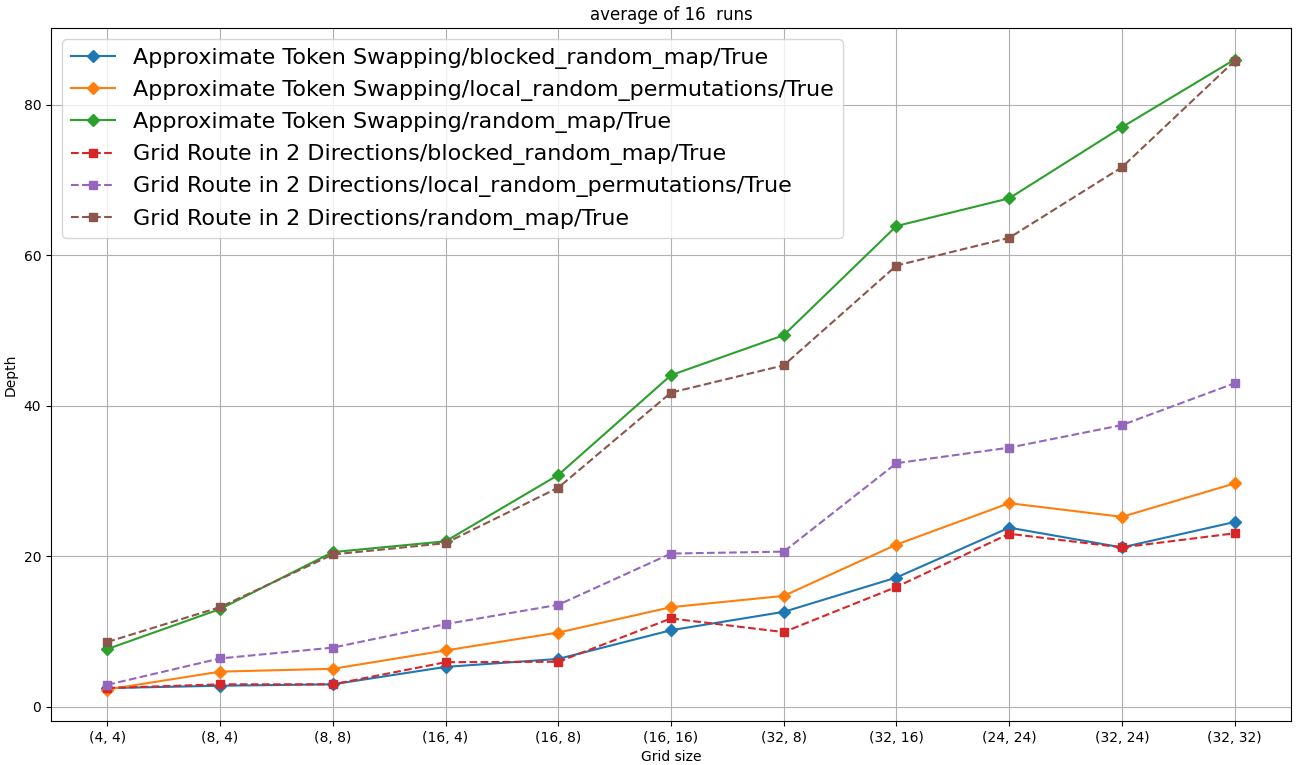}
  \caption{Depth of computed swap networks.}
  \label{depth}
  
  \includegraphics[width=\columnwidth]{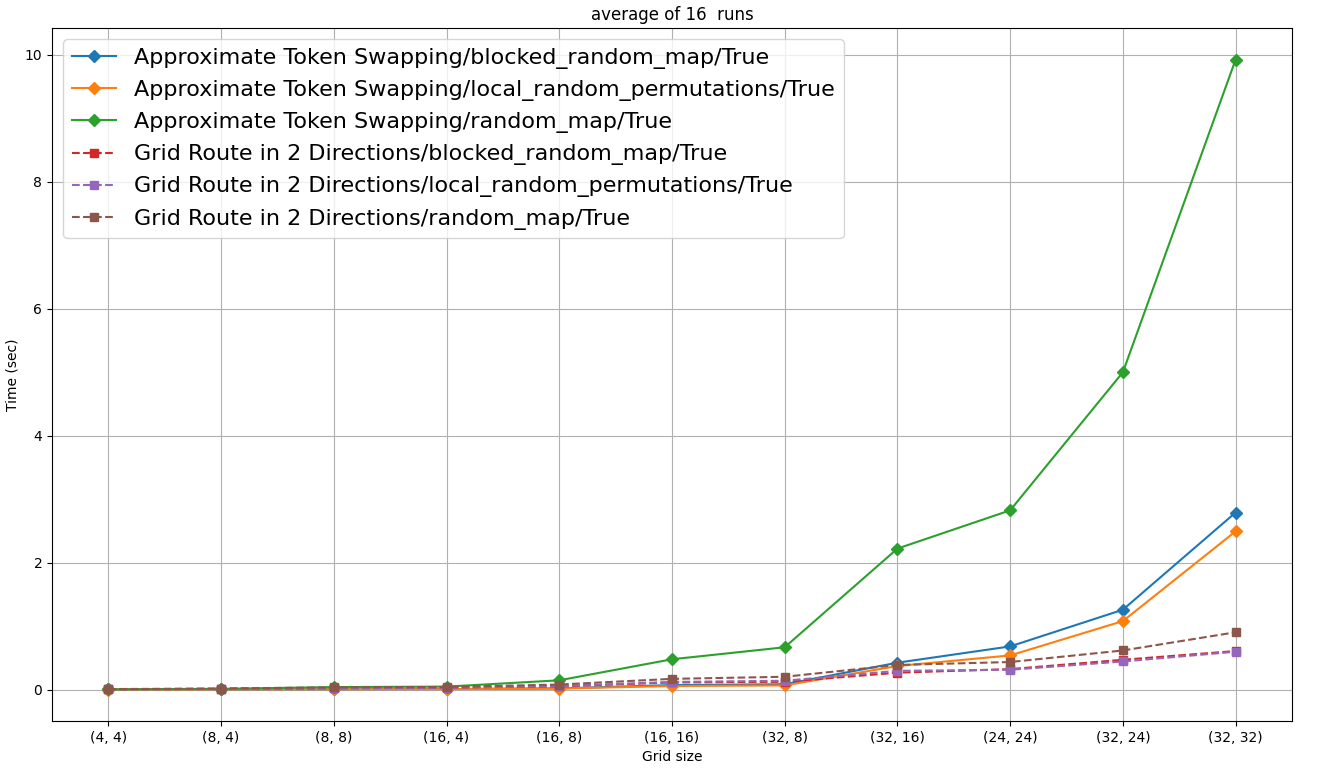}
  \caption{Time spent on finding swap networks.}
  \label{timing}
\end{figure}

Our locality-aware algorithm can always be made to produce a routing scheme with a smaller or equal depth as opposed to the naive grid routing algorithm. Otherwise, we can replace the output of the locality aware algorithm by that of the naive algorithm. This has virtually no computational overhead.
We compare our locality-aware grid router  against the  approximate token swapping (ATS) algorithm \cite{miltzow2016approximation} which has been used as a primitive on some state-of-the-art qubit transpilers (for example in \cite{childs2019circuit}). 
We set up the experiments based on a wide range of grid sizes and multiple random mapping schemes (local and global). Figures~\ref{depth} and ~\ref{timing}, respectively, summarize the effectiveness of the algorithm in terms of depth of the routing schedule and the execution time.
Figure \ref{depth} shows that our locality-aware router performs better than ATS when $\pi$ is a random permutation  (green vs brown plot in Figure \ref{depth}).
If the cycles of $\pi$ are constrained inside disjoint blocks then both algorithms seem to generate a routing schedule of similar depths (blue vs red plot in Figure \ref{depth}). 
On the other hand if the cycles of $\pi$ forms overlapping blocks, then ATS performs better than our algorithm.
If  $\pi$ happens to contain  long and skinny cycles that stretch in orthogonal directions, then our locality aware scheme will fail to optimize for both cycles simultaneously. This is not a bottleneck for ATS.
In terms of the running time we see that our algorithm scales well and in fact is significantly faster--an order of magnitude on larger grids vs ATS.
For our comparison we used the ATS implementation from \cite{childs2019circuit}. 
Our experimental data and source code can be found at \cite{repoGithub}.



\ifx false
Tests were carried out on the following grids:
(2, 2), (3, 2), (3, 3), (4, 4), (8, 8)

Current results show that local algorithm can find routing permutations up to
300\% and perform better as the size of the grid grows. The depth of the
permutations are comparable for smaller grid, but we expect the local algorithm
find shorter routings for larger grids.

(2, 2)
local: 0.001337486999999804
token swap: 0.0007112500000001631
speedup: 0.5317808696460357
(3, 2)
local: 0.002085516000000176
token swap: 0.0019294079999996328
speedup: 0.9251465824282671
(3, 3)
local: 0.0043778140000005905
token swap: 0.002477920999999661
speedup: 0.5660178801564724
(4, 4)
local: 0.0070019070000002515
token swap: 0.006919836000000679
speedup: 0.9882787646280407
(8, 8)
local: 0.030414485000000546
token swap: 0.09879344600000017
speedup: 3.248236687223157
(2, 2): depth approx = 1; depth localized = 1
(3, 2): depth approx = 4; depth localized = 3
(3, 3): depth approx = 5; depth localized = 6
(4, 4): depth approx = 5; depth localized = 9
(8, 8): depth approx = 21; depth localized = 21
\fi

\section{Conclusion}
In this extended abstract, we introduce an efficient routing algorithm for grid and Cartesian product architectures by taking advantage of the locality in the underlying permutation. Experiments demonstrate that the proposed method leads to comparable depth to a state-of-the-art algorithm with significantly higher performance. 

\bibliographystyle{IEEEtran}
\bibliography{ref}

\end{document}